\documentstyle[prl,aps,multicol,graphicx]{revtex}

\newcommand{\Ic}{$I_{\mathrm{c}}$}
\newcommand{\Tc}{$T_{\mathrm{c}}$}
\newcommand{\Jc}{$J_{\mathrm{c}}$}
\newcommand{\degree}{$^{\circ}$}
\newcommand{\scm}{cm$^2$}
\newcommand{\etal}{{\sl et al.}}

\begin{document}
\title{A Possible Solution of the Grain Boundary Problem for Applications of High-\Tc{}
Superconductors}

\author{G.~Hammerl, A.~Herrnberger, A.~Schmehl, A.~Weber, K.~Wiedenmann, C.\,W.~Schneider, and J.~Mannhart}

\address{Experimentalphysik VI, Center for Electronic Correlations and
Magnetism, Institute of Physics, Augsburg University, D-86135
Augsburg, Germany\\[12pt]}

\maketitle

\begin{abstract}
It is shown that the critical current density of high-\Tc{} wires
can be greatly enhanced by using a threefold approach, which
consists of grain alignment, doping, and optimization of the grain
architecture. According to model calculations, current densities
of $4\cdot10^6$\,A/\scm{} can be achieved for an average grain
alignment of 10\degree{} at 77\,K. Based on this approach, a road
to competitive high-\Tc{} cables is proposed.
\end{abstract}


\begin{multicols}{2}

\narrowtext

Vital for large scale applications of high-\Tc{} super\-conductors
\cite{bednorz,wu} is the solution of the grain boundary problem,
which manifests itself by the exponential decrease of the grain
boundary critical current density \Jc{} of the high-\Tc{} cuprates
as a function of the grain boundary
angle\cite{dimos,hilgenkamp:2002}.

We propose to solve this problem using a threefold approach:
through 1) grain alignment \cite{dimos}, 2) grain boundary doping
\cite{hammerl:2000}, and 3) optimization of the microstructure to
maximize the effective grain boundary area \cite{mannhart:1989b}.
In contrast to the powder-in-tube technology where large grain
boundary areas and grain alignment are used to enhance \Jc{}
\cite{mannhart:1990b,bulaevskii:1992,hensel:1995}, today's coated
conductor technologies \cite{iijima:1992,norton:1996,bauer:1999} focus on grain alignment
only. As we have shown, however, simple ways exist to also
preferentially dope the grain boundaries \cite{hammerl:2000} and
engineer large effective grain boundary areas
\cite{leitenmeier,hammerl:2002} to further enhance the performance
of coated conductors.

As pointed out in 1987, large effective grain boundary areas can
be realized by engineering the microstructure of the
superconductor to obtain grains with big aspect ratios, for
example by stacking in a brickwall-type manner platelet-like
grains on top of each other \cite{mannhart:1989b,mannhart:1990b}.
The enhancement of the critical currents hereby gained is
responsible for the large \Jc\ of the Bi-based high-\Tc{}
superconductors fabricated with the powder-in-tube technology
\cite{mannhart:1989b,mannhart:1990b,bulaevskii:1992,hensel:1995}.
Recently we found ways to use large effective grain boundary areas
to enhance \Jc{} of coated conductors that consist of two- or
three dimensional grain boundary networks, as illustrated in
Fig.\,1. Although it is clear that each one of the three
techniques described substantially enhances \Jc, the increase that
can be gained by utilizing all three, for example as shown by
Fig.~2, is unknown.

Therefore we have calculated the performance, which can be
achieved by combining grain orientation, doping, and large
effective grain boundary areas. Based on these calculations,
optimized sets of parameters for the fabrication of coated
conductors are derived.

\begin{figure}[b]
\centering\includegraphics[width=.7\columnwidth]{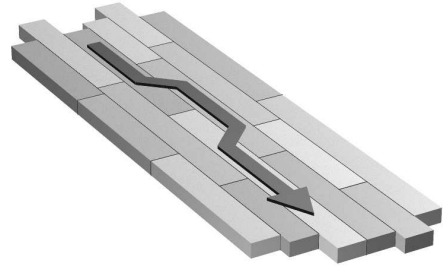}
\caption{Sketch of a coated conductor containing grains with big
aspect ratios. Large currents are supported by the conductor,
because bypasses around the standard, small area grain boundaries
are provided.}
\end{figure}

The calculation of current percolation through disordered networks
of weak links, some of which may be Josephson junctions, is a
complex problem \cite{mannhart:1989b}; and several algorithms have
been developed for its solution (see, e.\,g. Ref.~6, 15\,--\,19). As
the fast algorithms are limited to two-dimensional networks, for
the present work a new one had to be devised. Like in several of
the existing algorithms, to achieve the required speed, phase
effects and self fields were neglected.

\begin{figure}
\centering\includegraphics[width=.7\columnwidth]{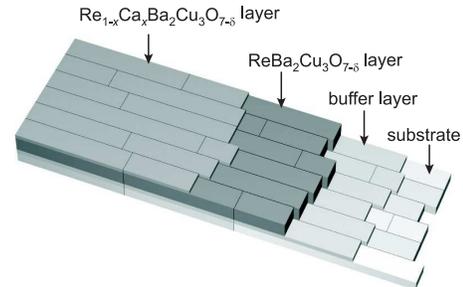}
\caption{Simplified sketch of a tape, fabricated by the rolling
assisted biaxially textured substrate (RABiTS) technology, with a
modified grain architecture which is based on grains with aspect
ratios $\rho\gg1$. Preferentially doping of the grain boundaries
is achieved by utilizing a doped cap layer and grain boundary
diffusion.}
\end{figure}

These calculations are based on Dijkstra's shortest path algorithm
for undirected graphs \cite{dijkstra}, the grains acting as
vertices, the grain boundaries, weighted by their critical
currents, as edges. To initialize a calculation, a polycrystalline
superconductor, typically containing  $10^3$\,--\,$10^4$ grains is
modelled first. To this superconductor an intragrain $J_{\mathrm
c,grain}$ and misorientation angle dependent grain boundary
critical current densities $J_{\mathrm{c}} =
J_{\mathrm{c}}(\theta)$ are ascribed. The calculations are
performed in steps, in each step~$i$ the algorithm finds in
Dijkstra's sense the shortest path through the network. The
critical current $I_i$ of this path and the respective current
densities are calculated and compared to the current densities of
the grains and boundaries that form this path. At the end of
step~$i$, the critical currents of the grains and grain boundaries
involved are reduced by $I_i$. After the final step $m$, all
possible current paths have been cancelled by this procedure and
the critical current $I$ is given by $I = \sum_{i=1}^m I_i$. These
calculations are repeated $N$ times to calculate \Jc{} for
different networks. The final result is obtained by averaging the
intermediate results.

Because the algorithm uses an undirected graph model, it is fast
and capable to determine the critical current of three dimensional
grain boundary networks. Its accuracy is a function of the grain
number and of $N$. The numerical accuracy of the data presented,
typically obtained with $10^3$ grains and $N = 20$, is better than
5\%.

This algorithm was used to assess possible approaches for the
optimization of coated conductors at 77\,K. For the present
calculations, the intragrain \Jc\ was taken to be
$5\cdot10^6$\,A/\scm, and the grain orientations and lengths were
chosen using Gaussian distributions with parameterized widths (see
also Ref.\,13). To determine the effects of doping on \Ic\, the
$J_{\mathrm{c}}(\theta)$-dependence in the simulation was modified
based on the experimental data \cite{hammerl:2000}, which show
that at least in the range of
$24^{\circ}\!\!<\!\theta\!<\!36^{\circ}$ the critical current
density can be doubled by doping. For lower and higher angles,
\Jc{} is taken to be exponentially reduced to the undoped values.

To consider the effects of stacking coated conductors in
multilayer configurations, the intergrain critical currents
flowing in c-direction were modelled by reducing
$J_{\mathrm{c}}(\theta)$ by an additional c-axis coupling factor
$f_{\mathrm{c}}$. For bilayers, the transverse misalignment of the
grains in the top and bottom layer was taken to be 30\% of the
grain width. These calculations provide a clear assessment of the
possibilities to optimize coated conductors, as shown in the
following.

In Fig.\,3 the critical current densities of various coated
conductors are plotted as function of the average grain
misorientation $\sigma$ and aspect ratio~$\rho$ of the grains. As
seen, the current density of conventional tapes ($\rho = 1$,
$\sigma \ge 15$\degree) is approximately doubled by doping the
grain boundaries, in agreement with experimental results. An
enhancement of the aspect ratio significantly increases \Ic\
further, and a tape with $\sigma=45$\degree{} and $\rho=50$ has
the same \Jc\ as conventional tape with an alignment of 6\degree.
This graph suggests to combine moderate grain alignment ($\sigma =
10$\,\degree), large aspect ratios ($\rho=20\mbox{\,to\,}30$), and
doping to achieve critical current densities of
$3-4\cdot10^6$\,A/\scm.

\begin{figure}
\centering\includegraphics[width=.75\columnwidth]{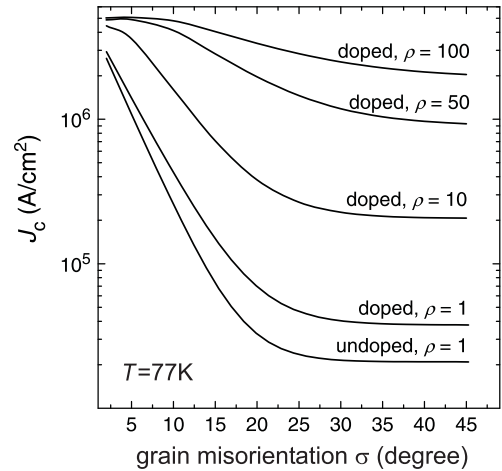}
\caption{Calculated critical current densities of various
\mbox{RABiTS} tapes with different grain aspect ratios $\rho$ as a
function of the grain misorientation $\sigma$.}
\end{figure}

As revealed by Fig.\,4, in which \Jc{} is plotted as a function of
aspect ratio $\rho$ and coupling $f_{\mathrm{c}}$ for two undoped
RABiTS tapes with 10$^{\circ}$ texturing stacked on top of each
other, the use of bilayers increases \Jc\ roughly by
$10^6$\,A/\scm. Bilayers therefore only provide a significant
advantage for conventional tapes with $\rho\approx1$. The benefits
are much smaller for tapes with larger aspect ratios and it seems
reasonable to apply for a given tape only either stacking or
elongation of the grains.

\begin{figure}
\centering\includegraphics[width=.75\columnwidth]{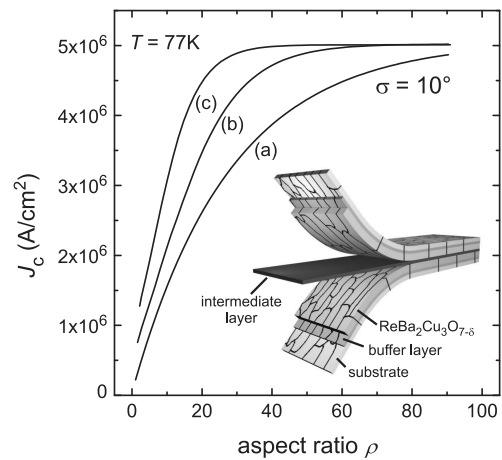}
\caption{Calculated dependence of the critical current density as
a function of the aspect ratio $\rho$ of two RABiTS tapes with
misorientation $\sigma=10^{\circ}$ stacked on top of each other
with different coupling factors $f_{\mathrm{c}}= 0$ (a), 10$^{-4}$
(b), and 10$^{-3}$ (c). The inset shows a sketch of such a tape,
the intermediate layer is used to weld the two tapes together.}
\end{figure}

The perspectives opened by the data shown in Fig.\,3 and Fig.\,4
suggest the mass production of high-\Tc{} tapes, as illustrated by
Fig.\,5.

\begin{figure}
\centering\includegraphics[width=.95\columnwidth]{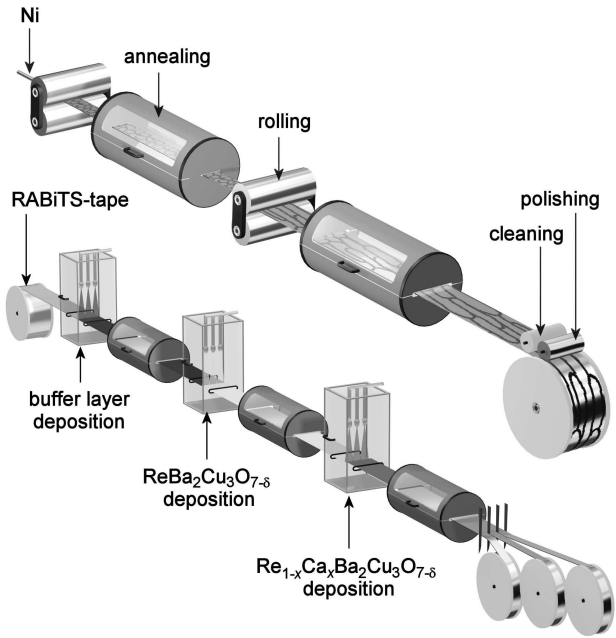}
\caption{Sketch illustrating large scale production of
ReBa$_2$Cu$_3$O$_{7-\delta}$ based RABiTS tapes as suggested by
the results shown in Fig.\,3 and 4.}
\end{figure}

This process is based on the standard RABiTS technology, and is
aimed to generate tapes with high critical currents by utilizing
grain boundary doping and tapes containing grains with big aspect
ratios. Here it is envisaged that Ni tapes with elongated grains
and proper texture can be fabricated, using processes as reported
in Ref.~21. A RABiTS tape is rolled and annealed to obtain an
average grain misorientation of $\approx 10$\degree\ and aspect
ratios in the range of 20\,--\,50. The buffer layer system and the
ReBa$_2$Cu$_3$O$_{7-\delta}$-film are deposited with non-vacuum
techniques \cite{araki}, before the tape is covered with a cap
layer of Re$_{1-x}$Ca$_x$Ba$_2$Cu$_3$O$_{7-\delta}$, annealed and
cut. According to Fig.\,3, such tapes support critical current
densities of $4\cdot10^6$\,A/\scm, corresponding to critical
currents of 400\,A/cm for 1\,$\mu$m thick, single sided
superconducting films. Although significant technological problems
associated with the growth of thick superconducting films,
ac-losses, and quench protection remain to be solved, the process
outlined appears to be suited for large scale production of coated
conductors with large critical currents.

We gratefully acknowledge helpful discussions with M.~Beasley,
J.\,G.~Bednorz, M.~Blamire, P.~Chaudhari, T.~Claeson, J.~Evetts,
H.~Hilgenkamp, B.~Holzapfel, Z.\,G.~Ivanov, S.~Leitenmeier,
D.\,G.~Schlom, and L.~Schultz. This work was supported by the BMBF
(13N6918).

\par
\end{multicols}

\begin{references}
\bibitem{bednorz}
J.\,G.~Bednorz and K.\,A.~M\"uller, Z. Phys. B {\bf65}, 189
(1986).

\bibitem{wu}
M.\,K.~Wu \etal, Phys. Rev. Lett. {\bf 58}, 908 (1987).

\bibitem{dimos}
D.~Dimos, P.~Chaudhari, and J.~Mannhart, Phys. Rev. B {\bf 41},
4038 (1990); Z.~G.~Ivanov \etal, Appl. Phys. Lett. {\bf 59}, 3030
(1991).

\bibitem{hilgenkamp:2002}
H.~Hilgenkamp and J.~Mannhart, Rev. Mod. Phys. {\bf74}, 485
(2002).

\bibitem{hammerl:2000}
G.~Hammerl \etal, Nature {\bf 407}, 162 (2000).

\bibitem{mannhart:1989b}
J.~Mannhart and C.\,C.~Tsuei, Z. Phys. B {\bf 77}, 53 (1989).

\bibitem{mannhart:1990b}
J.~Mannhart, "What limits the critical current density in
high-$T_{\mathrm{c}}$ superconductors?", in {\em Earlier and
Recent Aspects of Superconductivity}, edited by J.\,G.~Bednorz and
K.\,A.~M\"uller (Springer-Verlag, Heidelberg), 208 (1990).

\bibitem{bulaevskii:1992}
L.\,N.~Bulaevskii \etal, Phys. Rev. B {\bf 45}, 2545 (1992).

\bibitem{hensel:1995}
B.~Hensel, G.~Grasso, and R.~Fl\"ukiger, Phys. Rev. B. {\bf 51},
15456 (1995).

\bibitem{iijima:1992}
Y.~Iijima \etal, Appl. Phys. Lett. {\bf60}, 769 (1992).

\bibitem{norton:1996}
D.\,P.~Norton \etal, Science {\bf274}, 755 (1996).

\bibitem{bauer:1999}
M.~Bauer,
R.~Semerad, and H.~Kinder, IEEE Trans. Appl. Supercond. {\bf9},
1502 (1999).


\bibitem{leitenmeier}
S.~Leitenmeier \etal, Ann. Phys. (Leipzig) {\bf11}, 3 (2002).

\bibitem{hammerl:2002}
G.~Hammerl \etal, Eur. Phys. J. B {\bf27}, 299 (2002).

\bibitem{nichols:1991}
C.\,S.~Nichols and D.\,R.~Clarke, Acta Metallurgica {\bf39}, 995
(1991).

\bibitem{cai:1992}
Z.-X.~Cai and D.\,O.~Welch, Phys. Rev. B {\bf 45}, 2385 (1992).

\bibitem{hensel:1993}
B.~Hensel \etal, Physica C {\bf205}, 329 (1993).

\bibitem{rutter:2000}
N.\,A.~Rutter, B.\,A.~Glowacki, and J.\,E.~Evetts, Supercond. Sci.
Technol. {\bf 13}, L25 (2000); B.~Zeimetz \etal, Supercond. Sci.
Technol. {\bf14}, 672 (2001).

\bibitem{holzapfel:2001}
B.~Holzapfel \etal, IEEE Trans. Appl. Supercond. {\bf 11}, 3872
(2001).

\bibitem{dijkstra}
E.~Dijkstra, Numer. Math. {\bf 1}, 269  (1959).


\bibitem{mueller:1997}
F.\,E.\,H.~M\"uller, M.~Heilmaier, and L.~Schultz, Mat. Science
and Eng. {\bf234}, 509 (1997).

\bibitem{araki}
T.~Araki \etal, Supercond. Sci. Technol. {\bf 15}, L1 (2002);
Q.~Li \etal, Physica C {\bf357--360}, 987 (2001).

\end{references}
\end{document}